\documentstyle[prl,aps,epsf]{revtex}
\begin{document}
\draft

\twocolumn[\hsize\textwidth\columnwidth\hsize\csname %
@twocolumnfalse\endcsname

\title{ Variational Wavefunction for Quantum Antiferromagnets }
\author {  Franjo Franji\'{c} and Sandro Sorella}
\address {
 	Istituto Nazionale di Fisica della Materia,
  	International School for Advanced Studies,
	Via Beirut 2-4, I--34013 Trieste, Italy}  
\date{\today}
\maketitle
\begin{abstract}\widetext
We present here a new approach to determine an accurate variational 
wavefunction for general quantum antiferromagnets,
completely defined  by the requirement to reproduce the simple and well known 
spin-wave  expansion. By this wavefunction,
it is possible to obtain  the correct behavior of the 
long distance correlation functions  for the one dimensional $S=1/2$ 
antiferromagnet, i.e. {\em a system without long range order}. 
The  qualitative difference between  the integer and half integer case is also
easily understood with this variational approach. Finally we present numerical
results for the 2D XY model, showing that the present wavefunction 
has an overlap $ > 0.99 $  to the exact ground state of the 
$S=1/2$ model for finite system up to $6 \times 6$ clusters.
\end{abstract}
]
\narrowtext

Since the discovery of High-Tc superconductivity an increasing attention 
has been devoted to the study of strongly correlated systems. In particular, 
the role of antiferromagnetic (AF) correlations in 
such electronic systems was soon clear as it may lead to
superconductivity at finite doping.\cite{schrieffer} 

In the present paper we define a simple strategy to build a variational 
wavefunction for general quantum spin Hamiltonians, without restriction 
on dimensionality, spin, and existence or non existence of long range 
(AF) order. 

In order to simplify the discussion, we restrict the forthcoming analysis
to the general anisotropic $xxz$ model on a finite lattice with $L$ 
sites and with periodic boundary conditions: 
\begin{equation} \label{hamilt}
H= \sum\limits_{i,j} - J_{i,j} 
 (S^x_i S^x_j + S^y_i S^y_j ) + J^z_{i,j}  S^z_i S^z_j\, ,  
\end{equation}
where $\vec S$ is usual notation for spin $S$ operators and the 
couplings $J_{i,j},J_z$ depend only on the distance between the sites 
$i$ and $j$, in order to define a translation invariant Hamiltonian.
It is also assumed that the couplings $J$s allow a stable ferromagnetic
(FM) solution on the $xy$-easy plane, at the classical level.   
Within these notations, the spin isotropic Heisenberg model corresponds 
to nearest neighbor couplings with  $J^z_{i,j}=J_{i,j}$ for the 
AF case (after the usual transformation 
$S^+_i=(-1)^i S^+_i$ to change the sign of $J$ in a bipartite lattice)
and $J^z_{i,j}=-J_{i,j}$ for the FM case.   
The isotropic condition is obtained by continuity for $|J^z| \to J$.

For this general Hamiltonian (\ref{hamilt}) standard spin-wave (SW)
theory can be applied, where for convenience we set the order parameter
along the $y-$axis and apply Holstein-Primakoff transformation rules
at leading order in $1\over S$:
\begin{eqnarray}\label{spinwave}
S^y_i&=& -S + a^{\dag}_i a_i \ , \nonumber \\
S^x_i &=& \sqrt{S\over 2} (a^{\dag}_i + a_i )\ , \nonumber \\
S^z_i &=& i \sqrt{S\over 2} ( a^{\dag}_i -a_i). 
\end{eqnarray}
Here the boson $a^{\dag}_i$ creates a spin fluctuation at the site $i$
over the FM state $|F>$ 

By introducing standard Fourier transformed variables $a^{\dag}_q$,
$J(q)$ and $J^z(q)$, and after a little algebra, the linear SW  
Hamiltonian can be written in the compact form
$$H= S \sum_q \left[ D_q a^{\dag}_q a_q + {\eta_q \over 2 } 
( a^{\dag}_q a^{\dag}_{-q}  + {\it h.c.} ) \right], $$
where $D_q = 2 J(0)-J(q) + J^z(q) $ and $\eta_q = -(J(q) + J^z(q))$.
Each $q$ component of the above Hamiltonian can be dealt independently
and easily diagonalized by introducing the Bogoliubov transformation  
$a^{\dag}_q= u_q \beta^{\dag}_q +v_q \beta_{-q}$, with 
simple expressions for $u_q$ and $v_q$ in terms of $D_q$ 
and $\eta_q$.\cite{sorella}
In a finite lattice system, a special care should be payed to the
uniform $q=0$ mode corresponding  to the conservation of the total
spin component  along the $z-$axis. As shown in \cite{sorella}, 
the Bogoliubov transformation is singular for $q=0$. Nevertheless,  
this mode can be dealt systematically, because it is exactly equivalent
to a standard  projection of the SW  ground state to the subspace 
of vanishing total spin projection $S^z_{tot}=0$. Note that, in fact, the
classical state $|F>$ has not a definite $S^z_{tot}$ and after the
above projection, indicated in the following by $P_{S^z_{tot}=0}$,  
the order parameter is uniformly distributed in the $x-y$ plane, as is,
of course, expected in a finite lattice. 

In view of the above discussion, the ground state wavefunction has the 
following Gaussian form\cite{klr,sorella}:
\begin{equation}\label{wavefunction}
|\psi_G> =P_{S^z_{tot}=0}  \prod_{q\ne 0}  \left( u_q^{-1}
e^{\displaystyle{1\over 2}{v_q \over u_q} a^{\dag}_q a^{\dag}_{-q}}
\right) |F> .
\end{equation}
The SW result is very simple because represents the solution of a 
quantum oscillator for $S\to \infty$; however, in this limit the 
constraint, that the boson excitations on a single site obey 
$a^{\dag}_i a_i \le 2 S$, is obviously violated in the wavefunction 
(\ref{wavefunction}). 

We introduce here a simple wavefunction $|\psi_T>$ which is for any $S$ 
defined in the correct Hilbert space and only asymptotically
for $S\to \infty$ reduces to the known form (\ref{wavefunction}):
\begin{equation}\label{psit} 
|\psi_T> = P_{S^z_{tot}=0} e^{ \displaystyle  {1\over 2}
 \sum\limits_{ q  \in BZ} {2\over S} g_q S^z_q S^z_{-q} }  |F> . 
\end{equation}
This is a generalization of the Manousakis wavefunction
\cite{manousakis} obtained only for the isotropic AF model in 2D.
The unknown function $g_q$ is then determined by requiring that the
eigenvalue equation 
\begin{equation}\label{hpsi}
H |\psi_T> = E |\psi_T>
\end{equation} 
is satisfied for $S\to\infty$. Note that, due to the presence of the 
projector, the direction of the FM order parameter 
in the $x-y$ plane is irrelevant. It is thus convenient to direct it in
the $x-$axis direction before projection. In the state 
$|F>=\prod_i | S^x_i= S>$
each site has maximum spin along the $x-$axis, and in a basis diagonal
with $S_z|\sigma>=\sigma |\sigma>$, this state can be expanded as 
$|S^x_i=S>=2^{-S}\sum\limits_{m=-S}^S \sqrt{2S\choose S-m}|\sigma>$. 
It is therefore evident that the many-spin state $P_{S^z_{tot}=0}|F>$
can be generally written as a sum of classical configurations  
$|C_i> =|\sigma_1,\sigma_2,\cdots,\sigma_L>$, 
with $\sum_i\sigma_i=0$ with all positive, non vanishing coefficients.
For instance, for $S=1/2$,   
$P_{S^z_{tot}=0} |F>$ = sum all $S_z=0$ spin configurations $|C>$. 

The exponential form in (\ref{psit}) (commuting with $P_{S^z_{tot}}$) 
represents a classical Jastrow factor
$\exp\left({1\over 2}\sum_{i,j} v(i-j) \sigma_i \sigma_j\right)$
over the  possible  configurations $|C_i>$ with a proper two-body 
potential $v(r) = {2\over S L} \sum\limits_{q\ne 0} e^{-i q R} g_q $
acting on the classical spins.
The positiveness of the wavefunction over all such  configurations 
guarantees that (\ref{psit}) is not orthogonal to the true ground state,
for any finite $S$,  and therefore  it will necessarily
collapse to it, provided (\ref{hpsi}) is verified.\cite{lieb} 

We expect that the linear spin-wave approximation is quite accurate to
determine the self-consistency in the eigenvalue equation (\ref{hpsi}),
because this is determined only by the short range correlations, for
which there is always some kind of order, needed to apply the basic 
approximation (\ref{spinwave}). 
Provided  the physics of the ground state wavefunction 
is correctly described by (\ref{psit}), {\em also long distance 
spin-spin correlations} are expected to be correctly determined.
For instance, the spin-wave estimate is meaningless for the order
parameter $m=S -<a^{\dag} a> \to -\infty$ for the 1D $S=1/2$
Heisenberg model, whereas with the present approach the long range
order (LRO) of the wavefunction $P_{S^z_{tot}=0}|F>$ is readily destroyed
by the long range potential $v(r) \sim \ln r$, consistently determined
(see later) in the large $S$ limit.  
This approach allows therefore to define {\em a spin-wave expansion
even for models without a true LRO}, which represents a
remarkable extension  of this powerful technique for quantum spin
systems.\cite{zhong}  

In order to determine the function $g_q$, or equivalently to solve
(\ref{hpsi}) for $S \to \infty$, we just notice that each $q$ and $-q$
couple of wavevectors in (\ref{psit}) can be linearized by introducing 
complex auxiliary fields $z_q$:
\begin{equation}\label{hst} 
e^{ \displaystyle {2\over S} g_q S^z_q S^z_{-q} } =
 \int {d z_q\over \pi} e^{ \displaystyle 
-|z_q|^2 + \sqrt{{2g_q\over S}}\,( z_q S^z_q + z^*_{-q} S^z_{-q})}.
\end{equation}   
The above expression has to be applied to the vacuum of the spin waves 
$a_q|F>=0$. Since is, in this representation, 
$S^z_q= i \sqrt{S \over 2}  (a^{\dag}_q - a_{-q}) $, one easily
obtains that, for large spin, 
\begin{equation} \label{psils}
|\psi_T>  \propto P_{S^z_{tot}=0}  e^{ \displaystyle {1\over 2} 
\sum\limits_{q\ne 0}-{g_q\over 1 -g_q } a^{\dag}_q a^{\dag}_{-q} }|F> .
\end{equation}
By matching the two wavefunctions (\ref{psit}) and (\ref{wavefunction}), 
we determine $g_q=v_q/(v_q-u_q)$ for $S \to \infty$.  

\begin{figure}[m]
\epsfysize=8.75cm\epsfbox{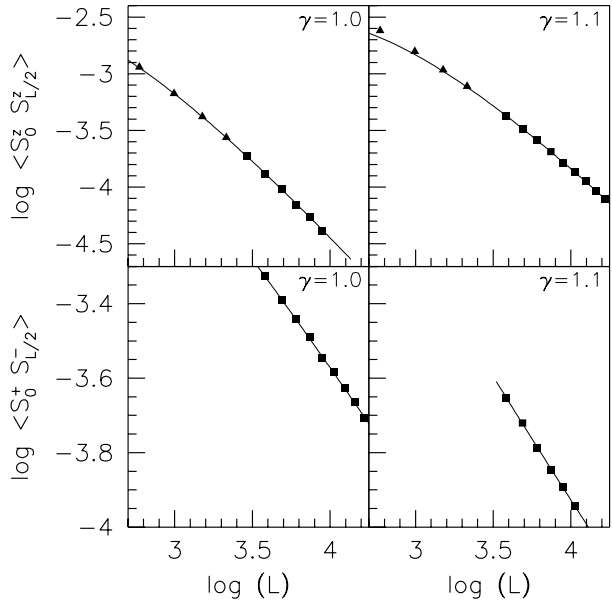}
\caption{
Log-log plot of the $S=1/2$ spin-spin correlation functions $c(L)$ for the
spin operators in a plane ($<S^+_0 S^-_{L/2}>$) and in the direction
orthogonal to it 
($<S^z_0 S^z_{L/2}>$) and for two values of $\gamma$. 
The triangles refer to exact diagonalization data and squares to Monte Carlo 
estimates (error bars are much smaller then size of the points). 
 The curve  are obtained by fitting 
the data to the power law behavior $\log c(L)=\mu\log L+a+b/L$:}
\null\hfill\begin{tabular}{|c|c|c|c|c|}
\hline
$\gamma$ & \multicolumn{2}{c|}{$<S^z_0 S^z_{L/2}>$} &
\multicolumn{2}{c|}{$<S^+_0 S^-_{L/2}>$}\\
\hline
 & $\mu_z$ & ${\pi\over 2\gamma}$ &
  $\mu_x$ & ${2\gamma\over\pi}$\\ 
\hline
$1.0$ & $1.59\pm 0.03$ & $1.5708$ & $0.64\pm 0.09$ & $0.6366$\\
\hline
$1.1$ & $1.43\pm 0.01$ & $1.4280$ & $0.6\pm 0.1$ & $0.7003$\\
\hline
\end{tabular}\hfill\null
\label{fig1}
\end{figure}

The function $g_q$, given by the above expression, turns singular only 
for $q\to0$ and behaving as $ \gamma_0 \over |q|$ with
$\gamma_0=-\sqrt{d}$ for the nearest neighbor model. By expanding
the exponential (\ref{psit}) in real space, we recover the wavefunction 
from Ref. \cite{mele} with logarithmic interaction between the spins.
As shown in this paper and related comments, {\em this is precisely the
condition to have Luttinger liquid behavior, or anomalous large distance
exponents in 1D}.
The spin-spin correlation functions, according to the Luttinger liquid
analysis, decay as a power low: $<S^z_0 S^z_r> \propto r^{-\mu_z} $ and
$<S^x_0 S^x_r > \propto r^{-\mu_x}$. Analogously to what was found in Ref.
\cite{mele}, we have obtained that $\mu_z$ and $\mu_x$ depend only on the
Jastrow coefficient $\gamma$: $\mu_z={\pi\over 2\gamma}$,
$\mu_x={1 \over \mu_z}$, in good agreement with the data shown in 
Fig. (\ref{fig1}).
For the isotropic model the prefactor is not the exact one which is
consistent with $<\vec S_0\cdot\vec S_r >\sim{1\over r}$ correlations.
However, if we determine $g_q$ by applying the SW expansion to the well
known Haldane-Shastry model\cite{haldane} ( $ J(q)= { 1\over 2 } q^2 $
and $J_z(q)= -{1\over 2} (\pi-q)^2 $ mod$(0,\pi)$), we find 
$\gamma_0= -{\pi\over 2}$, i.e. {\em the exact value}. The wavefunction
$|\psi_T>$ is the exact ground state of the model for particular 
$v(r)=2 \ln \sin({ \pi r \over L} )$ to which  our spin estimate 
is asymptotically converging for $L\to \infty$ 
( $g_q= -{\pi \over 2 |q|  } +\,\,\,  const.$).  

Let us now consider the spin one model, again in 1D.
In our approach, the function $g_q$ is insensitive to the spin $S$,
apart for a prefactor. However, the wavefunction {\em changes
dramatically}, compared to the $S=1/2$ case, due to the change of the
Hilbert space, determined not only by two spins with opposite magnitude  
($S_z=\pm 1$), but also by vacancy sites with $S_z=0$, that are
completely decoupled from the non-vanishing spins.
It is not difficult to realize that in the $S=1$ case the model
corresponds to an one dimensional Coulomb gas model with charge
$S_z=\pm 1$ since the vacancy sites can be considered as an empty space
between the charges. 

The Coulomb gas model with logarithmic interaction has been encountered
several times in the literature and its phase diagram, probably complete,
has been established.\cite{fisher}
But the mapping to this model is not orthodox, since with 
a little algebra the $\cos$ term in the Kane and Fisher Hamiltonian
is replaced by a $\log (\cos^{2S}( g \Phi))$ which differs from the 
usual model because there are infinite potential barriers between
the valleys. 
However, we have found that the
qualitative features of the $\cos$ model and the present $\log\cos^{2S}$
one are the same. 
Note also that the mapping to a model with a local defect
is possible  for integer $S$ and not for half integer spin, where 
our wavefunction displays  only one phase. 
\begin{figure}[m]
\epsfxsize=8.2cm\epsfbox{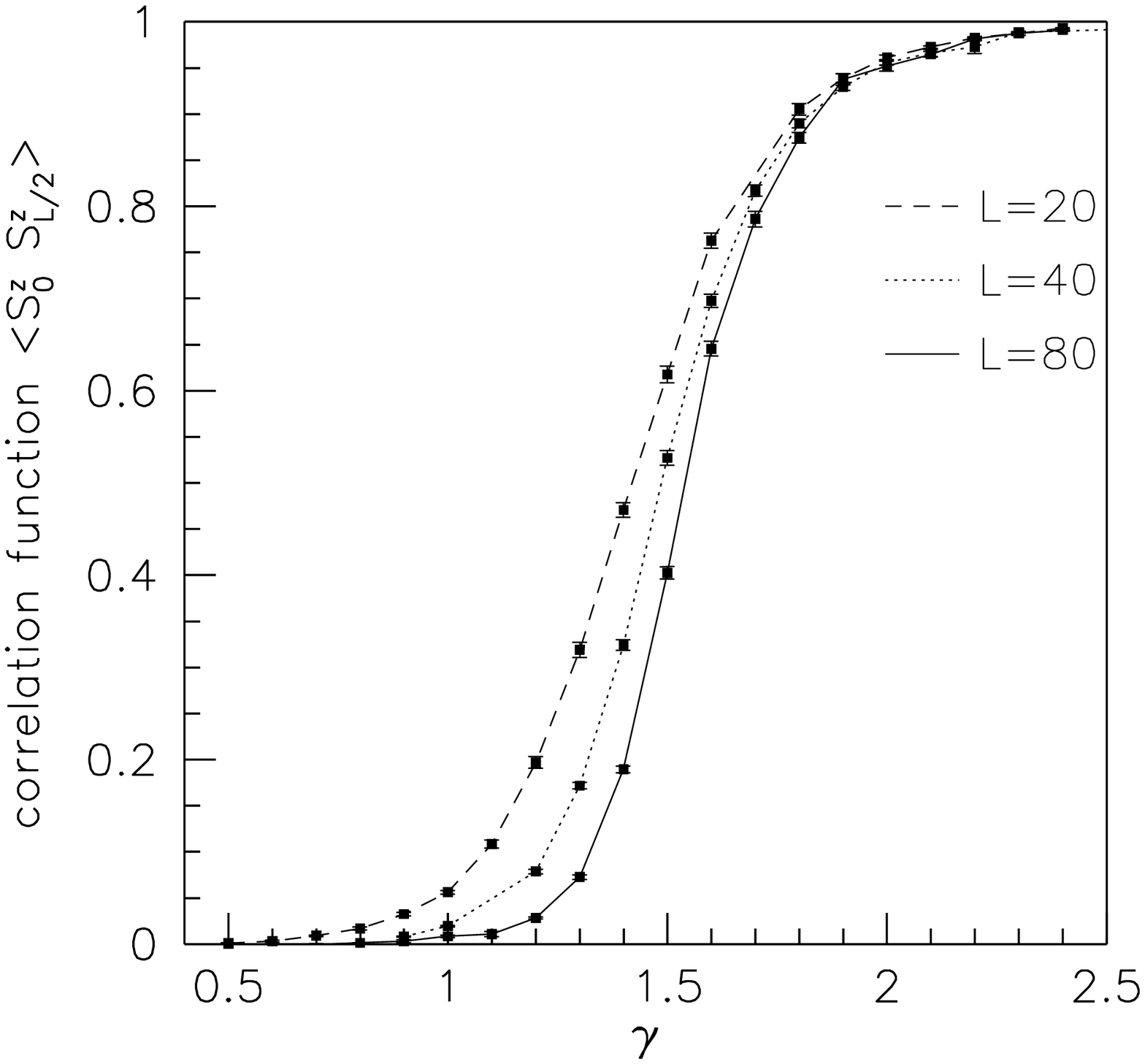}
\caption{
Plot of the $S=1$ spin-spin correlation function $<S^+_0 S^-_{L/2}>$,
for Monte Carlo data, as a function of $\gamma$.
When $\gamma$ increase, the system is going from a
disordered (plasma) phase, with a typical configuration\break
$0  +  +  -  +  -  0  0  0  0  +  -  +  0  0  0  -  0  0  +  -  0$,
to an ordered (dielectric) phase, with short range  
bound states with opposite spins \break
$-  +  -  +  -  +  -  +  -  +  -  +  -  +  -  +  -  +  -  +  -  +$.
The symbols  $+, 0, -$  represent the values ($S^z_i=+1,0,-1$).
For the larger system size, the change of the phase is more pronounced, 
implying a true phase transition as expected (see text).}
\label{fig2}
\null\vskip -0.5truecm
\epsfysize=8.75cm\epsfbox{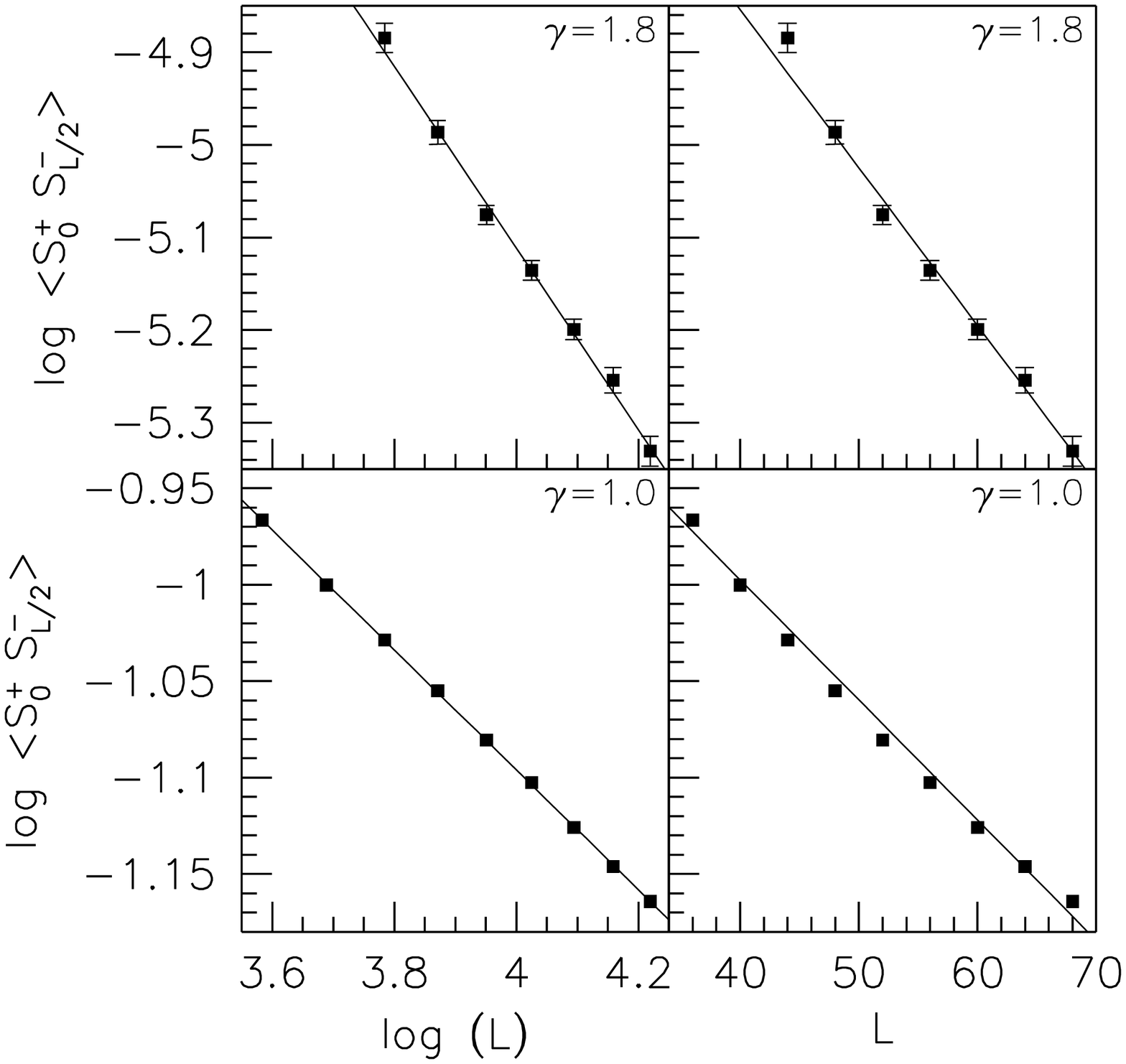}
\caption{
Log-log plots and L-log plots of the $S=1$ in-plane spin-spin 
correlation function $<S^+_0 S^-_{L/2}>$ for $\gamma=1.0$ and 
$\gamma=1.8$ (for the Monte Carlo data, with $\gamma=1.0$
 error bars are smaller then the size of points).
In the disordered phase ($\gamma=1.0$), the spin-spin correlation function 
clearly decays as a power low, whereas  the exponential decay assumption 
is much worse.}
\label{fig3}
\end{figure}

In the $S=1$ case,  for large $\gamma$,  there is a dielectric phase
(see Fig.\ref{fig2}) that reminds the Haldane phase 
with an hidden order parameter. Here the order is not hidden and of 
course the wavefunction is quantitatively meaningless. The remaining 
phase, where the charges are confined, is again characterized by a
power law decay in the spin-spin correlations. This suggests that, for
small $\gamma$, when our approximation is more reliable, a spin one
magnet has power law correlations (Fig. \ref{fig3}) and remains
gapless. Indeed, the phase diagram of the $S=1$ model\cite{nijs}
displays the power law  behavior close to the $J_z\simeq 0$ point,
which is consistent with our findings as  
the estimated $\gamma$  decreases for smaller  $J_z$. 
In the Kane and Fisher model, correlation
functions fall off with a power law in both regions. In the present model
we cannot exclude the existence of a correct finite correlation length
(gap) for integer spins, but we have not been able to find a clear
numerical evidence, as shown in Fig. (\ref{fig3}) for $\gamma=1.8$.
 Nevertheless, we
believe that the emerging picture in the integer spin case is
qualitatively sound and correctly describes the possible gapless phases,
present also in the $S=1$ models. 
\begin{table}[m]
\begin{tabular}{|c|c|c|c|c|}
\hline
Bosons &\multicolumn{2}{c|}{ $4 \times 4$} &
 \multicolumn{2}{c|}{$6 \times 6$} \\
\hline
 &$ E_V - E_{ex}$  &  $|<\psi_T|\psi_G>|^2$& $ E_V - E_{ex}$  & 
 $|<\psi_T|\psi_G>|^2$  \\
\hline
2&  0.00008 & 0.99997 & 0.00008 & 0.99994  \\
\hline
3 & 0.00040 & 0.99988 & 0.00014 & 0.99993  \\
\hline
4&  0.00128 & 0.99960 & 0.00034 & 0.99987  \\
\hline
5&  0.00236 & 0.99937 & 0.00071 & 0.99973  \\
\hline
6&  0.00415 & 0.99889 & 0.00126 & 0.99954  \\
\hline
7&  0.00642 & 0.99820 & 0.00198 & 0.99932  \\
\hline
8&  0.00792 & 0.99762 & 0.00288 & 0.99906  \\
\hline
9&          &         & 0.00396 & 0.99876  \\
\hline
10&         &         & 0.00524 & 0.99840  \\
\hline
11&         &         & 0.00674 & 0.99798  \\
\hline
12&         &         & 0.00842 & 0.99750  \\
\hline
\end{tabular}
\caption{ Difference between variational energy $E_V$
and exact one $E_{ex}$, and overlap square of the corresponding
wavefunctions for various clusters and number $N$ of bosons.
In this case the density of bosons was fixed by adding to the spin 
Hamiltonian (\protect\ref{hamilt}) 
a magnetic field term $-H  S^z_{tot} $ and determining the value 
of $H$ by requiring:
$ 2 S ({1\over 2}-\rho) = < S_z^{tot}> $ with $\rho= {N \over L}$ fixed 
and consistently $S \to \infty$. The direction of the order parameter has 
to be canted from the $xy$ plane in order to determine 
a stable classical solution. }
\end{table}

Finally, we want to show a simple application of this variational
wavefunction to a system of hard core bosons in the two dimensional case.
Provided the system has LRO for $\rho=N/L=1/2$ \cite{xyexact},
the spin-wave approximation is expected to be accurate. Still, it was 
amazing for us to see so much accuracy (see table) in {\em the ground
state wavefunction}, yielding practically {\em  an exact numerical 
solution of this correlated model}. Notice that, at fixed number of
bosons, the accuracy {\em improves } with size, showing that the low 
density limit\cite{abrikosov} is fulfilled exactly, by this wavefunction.
This is a non trivial fact because in this limit the Jastrow factor
diverges logarithmically. Our numerical results shows that the spin-wave
function (\ref{psit}) is very accurate for any density, especially in
the low density limit. We find therefore  a clear evidence 
that the superfluid condensate
exists, for any density, extending the rigorous proof in Ref. \cite{xyexact}
for $\rho=1/2$.

To conclude we have presented here a very promising method to  build 
unbiased variational wavefunctions for spin systems, which are quite 
reliable for gapless phases  because we reproduce in this case 
all the known results for  1D systems where the spin wave theory is worst. 
We believe that this wavefunction open {\em new} possibilities to understand 
strongly correlated systems.

One of us (SS) acknowledges a kind hospitality at the Cantoblanco University
in Madrid and useful discussions with F. Guinea.

\end{document}